\documentstyle[prl,aps,psfig,twocolumn]{revtex} 

\newcommand{\beq}{\begin{equation}}
\newcommand{\beqar}{\begin{eqnarray}}
\newcommand{\eeqar}{\end{eqnarray}}
\newcommand{\beqars}{\begin{eqnarray*}}
\newcommand{\eeqars}{\end{eqnarray*}}
\newcommand{\eeq}{\end{equation}}
\begin{document}
\draft
\twocolumn[\hsize\textwidth\columnwidth\hsize\csname
@twocolumnfalse\endcsname
\title{On the high dimensional Bak-Sneppen model}
\author{Paolo De Los Rios$^{(1)}$, 
Matteo Marsili$^{(1,2)}$ and
Michele Vendruscolo$^{(3)}$}
\address{$^{(1)}$Institut de Physique Th\'eorique, 
Universit\'e de Fribourg, CH-1700\\
$^{(2)}$International School for Advanced Studies
(SISSA) and INFM unit, ia Beirut 2-4, 34100 Trieste,
Italy\\
$^{(3)}$Department of Physics of Complex Systems, 
Weizmann Institute of Science, Rehovot 76100, Israel}
\date{\today}
\maketitle

\begin{abstract}
We report on extensive numerical
simulations on the Bak-Sneppen model in high
dimensions. We uncover a very rich behavior
as a function of dimensionality. For $d>2$
the avalanche cluster becomes fractal and
for $d\ge 4$ the process becomes transient.
Finally the exponents reach their mean field
values for $d=d_c=8$, which is then the upper
critical dimension of the Bak Sneppen model.
\end{abstract}
]
\narrowtext

The Bak-Sneppen (BS) model
\cite{bsmodel}, since its introduction, 
has attracted much attention in statistical
physics. Thanks to its simplicity, it has
lead to a much deeper understanding of the 
nature of self organized criticality\cite{PMB}
and of extremal dynamics \cite{RTS} in general.
The rules of the BS dynamics, in $d$ dimensions, are
very simple:  the state of the model is completely
defined by 
$L^d$ numbers $f_i$ arranged on a $d$-dimensional 
lattice of edge size $L$. 
At every time step the smallest of these numbers and 
its $2d$  
nearest neighbors are replaced with new uncorrelated 
random
numbers, drawn from the uniform distribution. 
Such very simple dynamics, based on the selection of
the global minimum, is generally called extremal
dynamics. It was first introduced in 
invasion percolation \cite{wilk} and it 
results  in a remarkably rich and interesting
critical behavior. 
 
The self-organized critical nature of the BS model 
(as well as of other extremal models) is revealed in 
its ability to naturally evolve towards a 
critical state where almost all of the variables
$f_i$ are above a threshold $f_c$. 
The dynamics in this state is characterized 
by scale-free bursts of activity or {\it avalanches}, 
which form a hierarchical structure 
\cite{bsmodel,backw}
of sub-avalanches within bigger avalanches.
This critical state is described by 
critical exponents. The distribution of 
avalanche duration $s$ behaves as a power law
$P(s)\sim s^{-\tau}$ with exponent $\tau$.
An avalanche of duration $s$ covers a number
of sites $V(s)\sim s^\mu$. The set of avalanche sites
is generally fractal $V(s)\sim R(s)^{D_f}$,
where $R(s)$ is the gyration radius 
and $D_f$ is the (spatial) fractal dimension. 
The active site (the one with the global minimum 
$f_i$) has a dynamical wandering that can be
described in terms of return times: 
The distribution $P_f(t)\sim t^{-\tau_f}$
of first return times is characterized by
an exponent $\tau_f$, whereas the distribution
of all return times $P_a(t)\sim t^{-\tau_a}$
defines the exponent $\tau_a$. For a random
walk these two exponents take the values
$\tau_f=3/2$ and $\tau_a=1/2$ in $d=1$,
whereas for $d\ge 2$ one finds 
$\tau_f=\tau_a=d/2$. Note that with respect
to previous literature\cite{bsmodel,PMB},
our notation is slightly different:
The exponent $\mu$ in ref. \cite{PMB} is 
defined as $\mu=d/D$, where 
$D$ was called ``fractal dimension''. 
Here we regard $\mu$ as a fundamental
exponent (not a composite one) and reserve
the name fractal dimension to describe the
{\em spatial} geometrical properties of the
avalanches. This choice, as we shall see,
avoids ambiguities which can arise in higher
dimensions, where the avalanche cluster
becomes fractal. 

A scaling theory was proposed in\cite{PMB} 
which shows that scaling relations allow
to reduce the number of critical exponents 
to two, for example $\mu$ and $\tau$.
Numerical simulations in $d=1$ and $2$
fully confirms the validity of the scaling 
theory\cite{PMB}. The mean field limit, 
formally corresponding to the limit $d\to\infty$,
has also been solved exactly \cite{mf}: the
exponents, in this limit take the values
$\tau=\tau_f=\tau_a=3/2$, $\mu=1$
and $D_f=4$. Finally, 
it was recently shown\cite{master,mdm} that a 
further non-trivial relation, of a different nature,
exists between $\mu$ and $\tau$. This suggests
that the BS universality class is characterized
by a single exponent, e.g. $\mu(d)$, as a function 
of $d$. 

In this Letter we analyze the behavior of the
BS model as a function of dimensionality. The
understanding of the critical behavior of a model, 
as a function of dimensionality, is a central
issue in statistical physics. In particular the
identification of the upper critical dimension $d_c$,
above which the mean field picture applies, is of 
great importance. Indeed it allows to understand
the behavior of a finite dimensional system using the
powerful tools of dimensional ($\epsilon$) expansion.
In equilibrium statistical mechanics, this is almost
routine work, but for non-equilibrium systems it is 
a challenging issue. For this reason the understanding
of the behavior of simple models as a function of
dimensionality is of great importance. 

We present extensive numerical
simulations for the BS model which show that 
{\em i)} the upper critical dimension is $d_c=8$ 
{\em ii)} for $d>2$ the avalanches
are no more compact $D_f<d$ 
{\em iii)} in $d=3$ we find $\mu<\tau_a<1$ and 
{\em iv)} for $d\ge 4$ the process becomes transient
i.e. $\tau_a=\tau_f>1$.
The BS model then shows a quite rich behavior, 
with four qualitatively different regimes
($d\le 2$, $2<d<4$, $4\le d<8$ and $d\ge 8$), 
as a function of dimensionality.
In particular we find that the relation 
$\mu=\tau_a$ holds only up to $d=2$ (note that this 
relation is in any case problematic close to the
mean field limit $\mu=1,~\tau_a=3/2$). 
A possible interpretation of our results is that 
``geometric'' exponents, such as $\tau_a,~\tau_f$ and
$D_f$ become independent from ``avalanche'' or
``memory'' (see later) exponents for $d>2$ and the
number of independent exponents changes with dimension. 
We shall first discuss in detail the numerical
procedure, defining operationally the quantities we
measure. Then we present the numerical results
and finally we discuss them.

The BS dynamics can be simulated very efficiently 
using a tree like search and replace
algorithm\cite{grass}. This decreases greatly
computation times, so that memory is the only
limitation to the system sizes studied. 
The numerical procedure was first tested in 
$d=1$ and $d=2$ and we found complete agreement
with refs. \cite{PMB,grass}. In order to compute
the exponents $\mu$ and $\tau$, following ref.
\cite{memo}, we introduce an {\em age} variable $k_i$
on each site.
The age $k_i$ of site $i$
measures the time elapsed since the last
update of the variable $f_i$. 
This method enables us to give a 
precise evaluation of the exponent $\mu$. Indeed
at each time the sites with age $k_i$ 
less than $s$ identify the current avalanche of
duration $s$. Therefore $V(s)$ is simply obtained
counting the sites with $k_i< s$. 
Age variables also allow 
for a determination of the avalanche exponent $\tau$. 
In systems evolving by an extremal dynamics,
it has been found\cite{RTS,memo} that the
probability that the global minimum $f_i$ occurs
on a site with age $k_i$ behaves as 
$\rho(k_i)\sim k_i^{-\alpha}$. 
The exponent $\alpha > 0$ implies that older sites are less 
likely to be selected. This reflects the fact that a very old
site has survived many selections, and therefore
it has very likely a high value of the variable $f_i$:
The more it survives the higher is its $f_i$, possibly 
even greater than the threshold $f_c$, in which case the site will
never be selected before a new update (occurring if one of its
neighbors is selected).
The exponent $\alpha$ is
related to the avalanche exponent by $\alpha=3-\tau$
\cite{alpha}. The advantage of measuring 
$\alpha$, with respect to a direct measure of
$\tau$, is that the statistics of the former is
much richer than that of the latter in the same
simulation. We measured $\tau$ in both
ways and found good agreement (which also
supports the validity of the relation 
$\alpha=3-\tau$). Since statistical uncertainty of
the exponent $\alpha$ is much less than that
of $\tau$ we report here only the value
$\tau=3-\alpha$. The return times exponents are
measured in the usual way\cite{PMB}: let
$t_i^{(k)}$, $k=0,1,\ldots$ be the (return) times
when site $i$ is visited ($t_i^{(k)}<t_i^{(k+1)}$).
The first return exponent is obtained
from the statistics of $t_i^{(k)}-t_i^{(k-1)}$,
whereas the all returns exponent is obtained
from $t_i^{(k)}-t_i^{(0)}$.
Finally, in order to obtain the fractal dimension
$D_f$, we compute the gyration radius of avalanches
of size $V(s)$.

The numerical results are summarized in Table \ref{tab1}. 
As shown in Fig.\ref{tau_mu}, our numerical
results for $\mu$ and $\tau$ are in good agreement
with the exact relation recently found in ref.
\cite{mdm}. This is an important consistency check
for the reliability of the simulations.
In $d=2$ we find a slight difference 
between $\tau_a$ and $\mu$ which is however 
within errorbars. The data are also plotted in 
Fig.\ref{tau_all_mu} for completeness. For $d=3$ 
the difference between $\tau_a$ and $\mu$ is much
larger than the error bars (see Fig.\ref{tau_all_mu}).
In Fig.\ref{tot} we plot the exponents as a function of dimensionality.
For $d\le 3$ the process is recurrent, since
$\tau_a<1$, i.e. each visited site, in an infinite system, is visited 
again an
infinite number of times (or, stated differently, a selected
site will be selected again with probability one).
Instead for $d\ge 4$ the process becomes transient
i.e. $\tau_a=\tau_f>1$: each site, in an infinite
system, is visited a finite number of times (there is a finite, 
smaller than one,
probability that a selected site will be selected again).
The relation $\tau_a + \tau_f = 2$ if $\tau_a<1$
and $\tau_a = \tau_f$ if $\tau_a \ge 1$ 
(a classical result from renewal theory\cite{fisher})
is always respected, the former holding
for $d\le 3$. 
The return time statistics, in the BS model, 
is determined both by memory effects and by
geometry. Activity can return to site $i$
either because all sites with variables
larger than $f_i$ are eliminated by the BS 
dynamics (which we call a memory return) 
or because the activity returns close to site $i$ and
$f_i\to f_i'$ is  updated (geometric return). 
As the dimension $d$ increases, geometric
returns become less and less relevant (see later) and
for very large dimensions one expects to recover
the mean field result $\tau_a=\tau_f=3/2$.
We find that the mean field limit holds
for $d\ge d_c=8$. 

We now turn to the discussion of the onset of
fractality of avalanche clusters for $d > 2$.
At $d=2$, we found numerically that
the fractal dimension is slightly smaller than
$D_f=2$. We argue, however, that small size effects
occur and that $D_f=2$ holds.
Indeed, as the system size $L$
is increased the slope of the curve $\log V(s)$
vs $\log R(s)$ increases, suggesting that 
avalanches are compact. The occurrence
of small size effects can be understood 
analyzing the dynamics of the growth of the
avalanche cluster in $d$ dimensions. 
The growth probability distribution of the cluster 
is characterized by memory. 
The problem of cluster growth with memory was
recently investigated in ref. \cite{grwm}.
In a standard model of growth, new sites can be added
only at the surface of the cluster.
In this case, it was shown that when $\alpha>1$ 
the cluster has a fractal dimension $D_f<d$ \cite{grwm}. 
The difference between cluster growth
with memory and the 
avalanche cluster growth in the BS is that
in the latter the activity can return to bulk sites. 
At the initial stages of growth of BS clusters, however,
the analogy is useful, since  the ratio between
the number of surface sites and that of bulk 
sites is $ \sim 2d$ so that returns to the bulk are
rare. In $d=2$, as more and more returns to the bulk 
occur the surface to volume ratio decreases
and the cluster becomes compact (for a fractal 
cluster the ratio of  perimeter sites to 
bulk sites is finite). Therefore, while
small $s$ avalanches look fractal, for
larger sizes they become more and more
compact. These small size effects are
also visible in the distribution of
first return times (see also \cite{PMB}).
In $d\ge3$ the same size effect should appear,
however in this case we find a fractal dimension 
which is definitely smaller than $d$ \cite{notadf}.
Returns to the bulk are 
fewer in $d\ge3$ than in $d=2$ and more
importantly the topology of a fractal
cluster is very different: while the 
$2$-dimensional cluster is characterized by a
distribution of holes of all sizes, a fractal
in $d\ge3$ is most likely a ramified object.
Returns to the bulk can fill the holes
of the $d=2$ cluster 
but it is much more difficult for them
to turn the branched fractal structure of the 
$d\ge3$ avalanche into a compact one.
For this reason, we believe that 
our data are compatible with
the occurrence of a compact
cluster in $d=2$ and 
with $D_f<d$ when $d\ge3$. 
In any case, numerical data are relatively
stable with changes of the system size and
we could not detect any small size effect 
such as the one discussed above for $d=2$.

Geometric returns arise because the avalanche 
cluster has many self-intersections: indeed the
self-intersection  set has a fractal
dimension $D_I=2D_f-d>0$ for
$d<8$\cite{codim}. Note that $D_I$ is smaller than
$D_f$ which means that
the larger the avalanche the smaller is
the fraction of intersection sites. Moreover,
as mentioned earlier, geometric returns become
less and less relevant as $d$ increases
since $D_f-D_I=d-D_f$ also increases with $d$.
Above $d=d_c=8$ the fractal dimension
$D_f=4$ attains its mean field value and the 
avalanche has no self intersections ($D_I<0$).

The fractality of avalanches implies that some
scaling relations must be modified. 
For example the jump probability 
distribution $\rho(r)$ was analyzed in ref. \cite{PMB}
under the hypothesis of a compact cluster.
It is defined as the 
probability that the activity jumps in one
time step to a site at a distance $r$ from
the current site and it falls off with a 
power law behavior: $\rho(r)\sim r^{-\pi}$.
The exponent $\pi$ was related to $\tau$ and
$D$ in ref. \cite{PMB} by $\pi=1+D(2-\tau)$,
under the hypothesis of a compact cluster.
The same argument as in ref. \cite{PMB}
led us to the expression 
\beq
\pi=1+\frac{D_f(2-\tau)}{\mu}.
\eeq
It is interesting to observe that, for the values
quoted in Table \ref{tab1}, $\pi$ is always
slightly larger than $3$, so that the second
moment of $\rho(r)$ is finite, and its
mean field value is $\pi=3$. Were it not for
the correlations, the activity would perform
a random walk and it would be transient already
for $d>2$. Correlations induced by memory effects,
shifts the onset of transient behavior to
$d\ge 4$.

In order to check the full consistency of the
results shown in Table \ref{tab1}, we are also exploring
an alternative way to move away from the mean field limit
\cite{DLR}.
Within a $d=1$ system, we choose the ``nearest neighbors''
of the active site at random over the lattice, 
with a probability which is a power law decreasing
function of the distance from the active site, with
exponent $\omega$. Preliminary results show
that the mean field limit is recovered when $\omega \to 1$.
Varying $\omega > 1$ again three different critical regimes
appear. In particular we find a region of the values of $\omega$
where activity is recurrent ($\tau_a <1$)
but $\mu \ne \tau_a$ and a region where activity is transient ($\tau_a > 1$)
but $\mu < 1$. Moreover we find that, whenever $\mu$ takes the same
values listed in Table \ref{tab1}, also the other
quantities take on the same corresponding values (apart of
course from $D_f$ which is always smaller than $1$).
Finally, working in $d=1$ allows us to simulate extremely large sizes and 
to rule out any finite size effects,
thus strenghtening the reliability of the results shown in Table \ref{tab1}.

It has been argued that the mean field limit
of BS can be described as a branching 
process\cite{mf}. The value $d_c=8$, which coincides 
with the upper critical dimension of branched
polymers, is consistent with this picture\cite{flory}.
We believe that close to $d=8$ branched polymers
give a reasonable description of the geometry of
the BS process. An $\varepsilon=d_c-d$
expansion for the BS model could therefore
be feasible, also using the recent expansion
around the mean field solution of ref. \cite{mdm}.
In this respect, our results allow for a prediction
of the first coefficient $\mu\simeq 1-0.017\varepsilon$
of the $\varepsilon$ expansion.

In summary, we have presented numerical results
for the BS model in high dimensions. These allow
to conclude that for $d>2$ avalanches become fractal
and for $d\ge 4$ the process becomes transient.
Finally the mean field limit is reached at the
upper critical dimension $d_c=8$.
This behavior, with three different non-trivial
regimes of criticality, is substantially richer that 
that of equilibrium statistical models, where
non-trivial behavior occurs only in one region
of dimensionality, limited by the lower and the upper
critical dimension. Our results also call for
a close analysis of the scaling theory 
of ref. \cite{PMB}, which has been developed 
under the implicit assumptions of a recurrent
process with compact avalanches.

\begin{table}[ht]
\begin{center}
\begin{tabular}{c|cccccccc}
$d$              & 1    & 2    & 3    & 4    & 5    & 6    & 7     & 8    \\
\hline
$\mu$            & 0.42 & 0.69 & 0.85 & 0.92 & 0.95 & 0.96 & 0.98 & 1.00  \\
$\tau$           & 1.07 & 1.25 & 1.35 & 1.41 & 1.45 & 1.46 & 1.48 & 1.50 \\
$\tau_a$         & 0.42 & 0.70 & 0.92 & 1.15 & 1.29 & 1.40 & 1.49 & 1.50  \\
$\tau_f$         & 1.58 & 1.28 & 1.09 & 1.16 & 1.29 & 1.40 & 1.49 & 1.50  \\
$D_f$            & 1    & 2    & 2.6  & 3.3  & -    & -    & -    &  -  \\
$n$              & 21   & 10   &   7  & 5    & 4    & 3    & 3    & 2   \\
\end{tabular}
\caption{Exponents of the BS model for different dimensions $d$. 
An upper bound on the error is $\pm 0.01$ for safety, even if
for some of the listed exponents confidence is greater. In particular,
for $d=7$, the values are distingushable from
the $d=8$ ones. The last line
gives the size $L=2^n$ of the edge of the hypercube 
used for the simulations. }
\label{tab1}
\end{center}
\end{table}

\begin{figure}
\caption{Avalanche exponent $\tau$ vs. $\mu$ for different dimensions
(crosses). The continuous line is the exact relation
between the two exponents as from ref.[7]. The data from the simulations
and the exact relation are in excellent agreement.}
\label{tau_mu}
\end{figure}

\begin{figure}
\caption{Log-log plot of the number of covered sites $V(t)$ (dashed line) 
and the (inverse) all return times  
distribution $P_a(t)$ (full line) as a function of time $t$. 
The results in a) $d=3$
show that the slope of the two quantites are clearly different (hence
$\mu \ne \tau_a$) b) $d=2$ show instead full compatibility of the 
slopes with $\mu = \tau_a$. }
\label{tau_all_mu}
\end{figure}

\begin{figure}
\caption{The exponents $\mu$, $\tau$, $\tau_a$ and $\tau_f$ 
as a function of dimensionality. Dashed lines at $1$ and $1.5$ have been drawn
for reference.}
\label{tot}
\end{figure}


\begin{references}
\bibitem{notadf}
The measurement of the fractal dimension
is the most time consuming and moreover the 
feasible system sizes are smaller the larger 
the dimension. For this reason we did not
push the calculation of $D_f$ to $d>4$.
Note that the mean field result $D_f=4$ implies
$D_f\le 4$ and therefore a fractal structure
for $d\ge 5$.
\bibitem{codim}
The dimension $D_I$ of the intersection of
two objects of fractal dimension $D_f$ in $d$ dimensions 
can be obtained knowing that
the codimension of the intersection ($d-D_I$) is the sum of the codimensions
of the two intersecting objects ($d-D_f$). Therefore $d-D_I = 2 (d-D_f)$
from which the relation $D_I = 2 D_f -d$ follows.
\end{references}
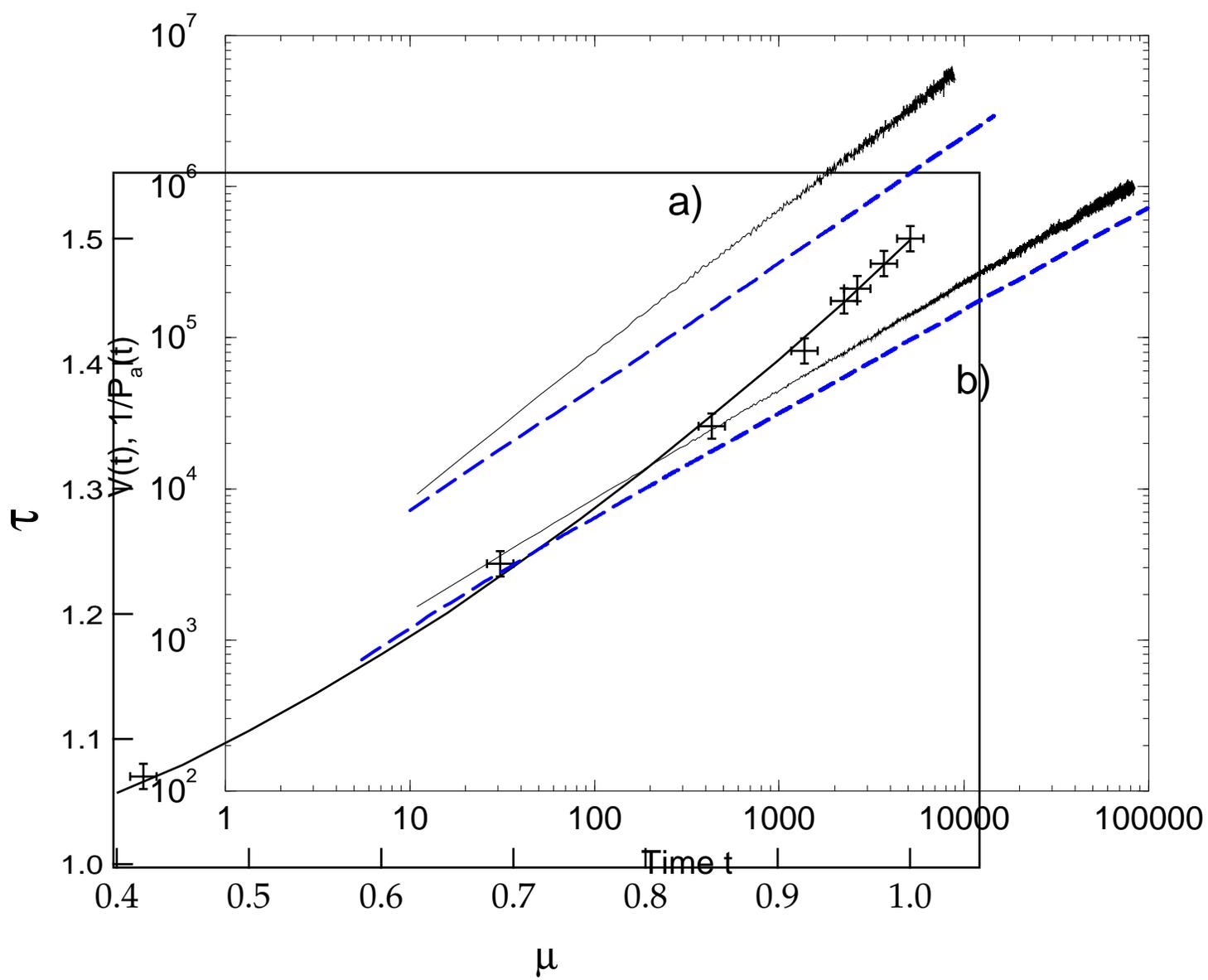
\end{document}